\documentclass[
 reprint,
 aps, superscriptaddress
]{revtex4-1}

\usepackage{graphicx}
\usepackage{dcolumn}
\usepackage{bm}
\usepackage{color}

\begin{document}

\preprint{APS/123-QED}

\title[Sample title]{Optomechanical magnetometry with a macroscopic resonator}


\author{Changqiu Yu}
\affiliation {National Key Laboratory of Tunable Laser Technology, Institute of Opto-Electronics, Harbin Institute of Technology, Harbin 150080, China.}
\affiliation{School of Mathematics and Physics, University of Queensland, Brisbane, Queensland 4072, Australia.}
\author{Jiri Janousek}
\affiliation{Centre for Quantum Computation and Communication Technology, Department of Quantum Science, The Australia National University, Canberra, ACT 0020, Australia.}
\author{Eoin Sheridan}
\affiliation{School of Mathematics and Physics, University of Queensland, Brisbane, Queensland 4072, Australia.}
\author{David L. McAuslan}
\affiliation{School of Mathematics and Physics, University of Queensland, Brisbane, Queensland 4072, Australia.}
\author{Halina Rubinsztein-Dunlop}
\affiliation{School of Mathematics and Physics, University of Queensland, Brisbane, Queensland 4072, Australia.}
\author{Ping Koy Lam}
\affiliation{Centre for Quantum Computation and Communication Technology, Department of Quantum Science, The Australia National University, Canberra, ACT 0020, Australia.}
\author{Yundong Zhang}
\affiliation {National Key Laboratory of Tunable Laser Technology, Institute of Opto-Electronics, Harbin Institute of Technology, Harbin 150080, China.}
\author{Warwick P. Bowen}
\email{w.bowen@uq.edu.au}
\affiliation{School of Mathematics and Physics, University of Queensland, Brisbane, Queensland 4072, Australia.}

\date{\today}

\begin{abstract}
We demonstrate a centimeter-scale optomechanical magnetometer based on a crystalline whispering gallery mode resonator. The large size of the resonator allows high magnetic field sensitivity to be achieved in the hertz to kilohertz frequency range. A peak sensitivity of 131~pT~Hz$^{-1/2}$ is reported, in a magnetically unshielded non-cryogenic environment and using optical power levels beneath 100~$\mu$W. Femtotesla range sensitivity may be possible in future devices with further optimization of laser noise and the physical structure of the resonator, allowing applications in high-performance magnetometry. 
\end{abstract}

\maketitle

\section{Introduction}
Whispering gallery mode (WGM) resonators play an important role in modern optics, with applications as laser cavities~\cite{laser}, resonant filters~\cite{filter}, optical switches~\cite{switch}, and precision sensors~\cite{sensor1,sensor3,sensor4,sensor5} among other areas. They have been recently used for magnetometry~\cite{PRL,AM} based on the ideas of cavity optomechanics\cite{COM_review}. WGM resonator based optomechanical magnetometry combines the ultra-high optical transduction sensitivity of WGM resonators with the giant magnetostriction of materials such as Terfenol-D, achieving high sensitivity while allowing room-temperature operation and simple optical readout. These advantages may enable applications in areas such as geophysical surveying~\cite{geophysics}, tests of fundamental physics~\cite{funderm1,funderm2}, medical imaging~\cite{medical1,medical2}, and space exploration~\cite{space1,space2}.

Optomechanical magnetometers based on microscale on-chip WGM resonators have achieved 200 pT Hz$^{-1/2}$ magnetic field sensitivity at megahertz frequencies~\cite{AM,PRL}. However, due to a combination of noise sources at low frequency and poor low frequency mechanical response, magnetic field sensing in the hertz to kilohertz frequency range was only possible using  inherent mechanical nonlinearities within the magnetostrictive material. This indirect approach caused a sacrifice in sensitivity to 110 nT Hz$^{-1/2}$. The hertz-kilohertz frequency range is crucial to many applications including, for instance, magnetic anomaly detection \cite{MAD_paper}, geological surveying \cite{geo_survey_paper}  and magnetoencephalography \cite{MEG_paper}. To enable highly sensitive magnetic field sensing in this regime, we have developed a centimeter-scale crystalline WGM resonator based magnetometer, which features reduced thermomechanical noise, lower frequency mechanical resonances, and higher optical quality factor than previously demonstrated optomechanical magnetometers. By embedding the magnetostrictive material (Terfenol-D) within the WGM resonator, sub 10~nT~Hz$^{-1/2}$ sensitivity was achieved over most of the frequency band from 127~Hz to 600~kHz, with a peak sensitivity of 131~pT~Hz$^{-1/2}$ at 127~kHz.

\begin{figure}
\includegraphics[width=8.5cm]{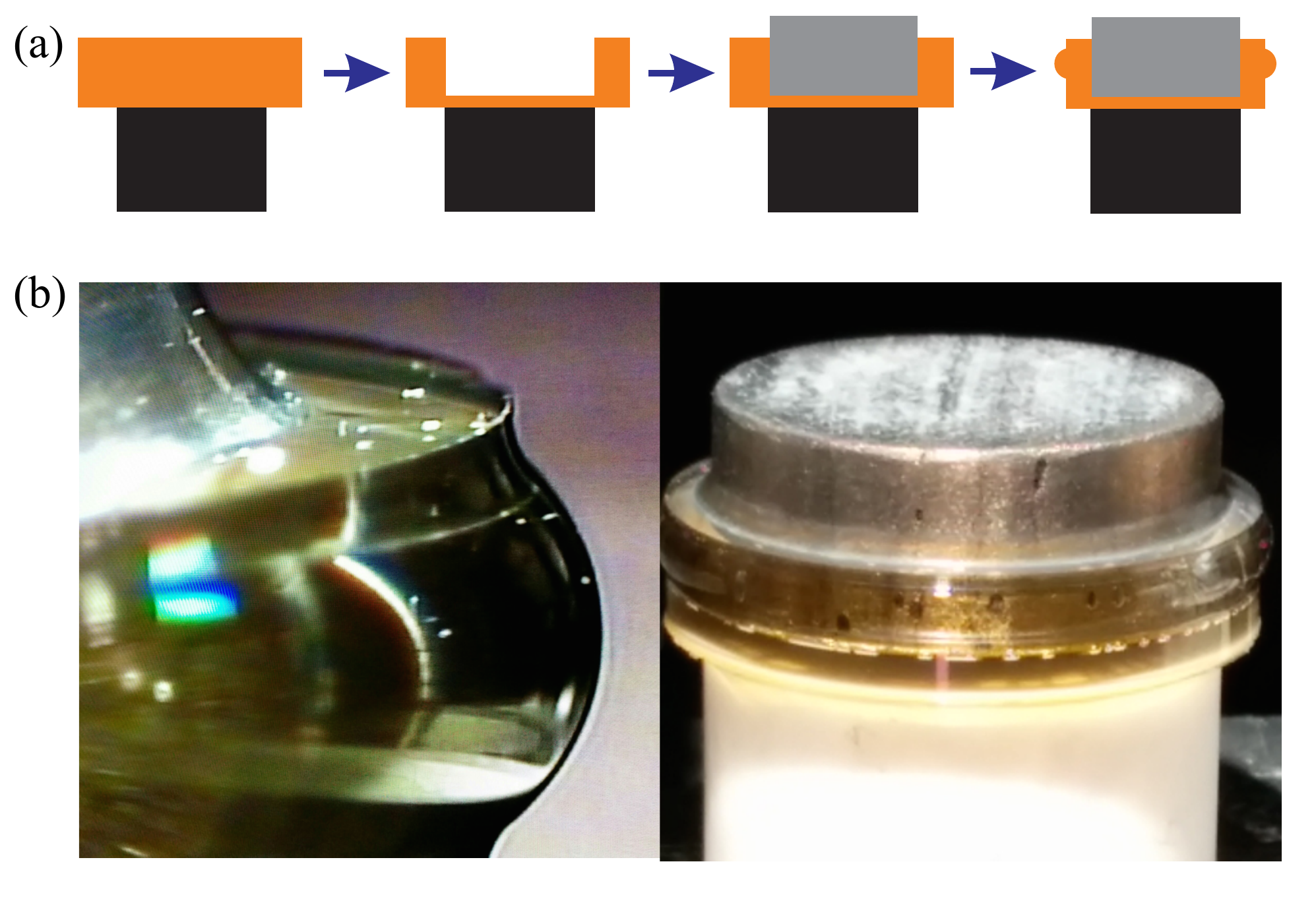}
\caption{(Color online) (a): The fabrication process. Black area: ceramic, yellow area: CaF$_2$, gray area: Terfenol-D [Etrema products Inc.]. (b) Optical microscope images of the resonator.}\label{resonator}
\end{figure}
\section{Resonator fabrication and characterisation}
The WGM resonator was fabricated using the Ultra-precision Machining Facility at the Australian National University, housing a Moore Nanotech 250 UPL diamond turning lathe. WGM resonators are particularly well-suited for fabrication by diamond turning due to their cylindrical symmetry. We fabricated the resonator from CaF$_2$ due, primarily, to the previously demonstrated capability to achieve exceptionally high optical quality factors using this material~\cite{experimenthighQ}. The fabrication process of the magnetometer is shown in Fig.~\ref{resonator}(a). A bulk of CaF$_2$ crystal, which was attached to a ceramic pedestal using a vacuum compatible epoxy glue (EPO-TEX 353ND), was first rough-cut to form a WGM resonator with a diameter of 16~mm. Lathing was also used to bore a void in the top of the crystal WGM structure. The void was machined to a diameter 30 $\mu$m larger than the actual size of the disk of Terfenol-D (of diameter and thickness approximately 12~mm and 4~mm, respectively). The 15~$\mu$m gap was the minimum that allowed the epoxy glue, due to its viscosity, to uniformly fill the interface of the two materials. Next, we machined the final WGM structure with the radius of curvature of the resonator's rim of 1.616~mm~\cite{Strekalov}.

The final step is to polish the resonator to achieve an extremely smooth surface, i.e., a high intrinsic optical quality factor. Using the lathe to rotate the WGM resonator and ensuring that the resonator is precisely centred on the rotational axis, polishing was accomplished using a polishing pad and diamond slurry. Starting with 0.5~$\mu$m particle size, large chips on the surface of the resonator left after cutting were removed and using progressively smaller particle sizes down to 0.05~$\mu$m, the final polishing was achieved. The physical structure of the resonator is shown in Fig.~\ref{resonator}(b).

\begin{figure}
\centering
\includegraphics[width=8cm]{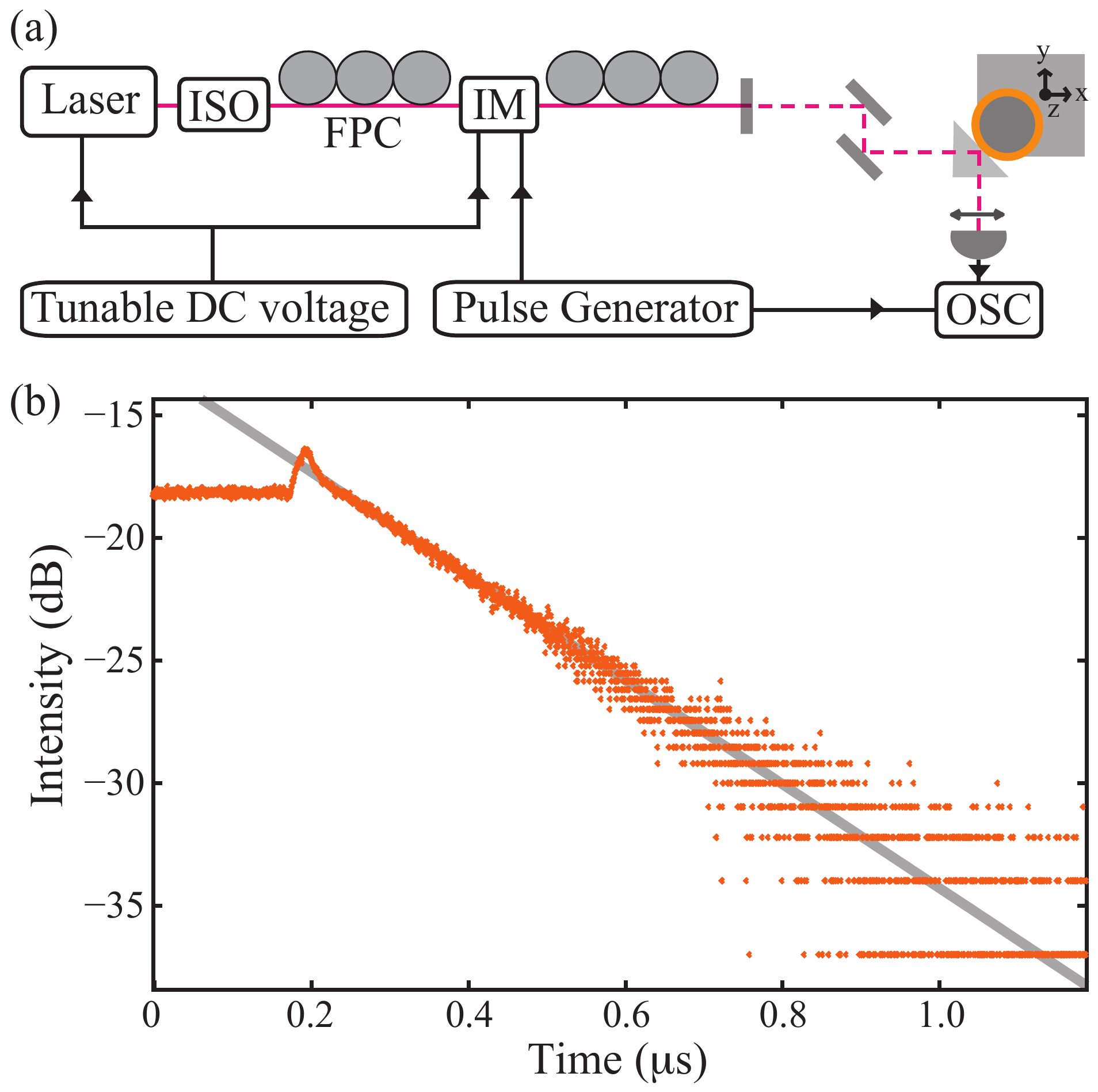}
\caption{(Color online) Ringdown measurements. (a): Schematic of the apparatus used to perform ringdown measurement of the optical quality factor.  FPC: fiber polarization controller, IM: intensity modulator [OC-192 Modulator JDS Uniphase], OSC: oscilloscope [Tektronix TDS 2024B], pulse generator [Stanford DG535], nanomax stage [Thorlabs MDT693A], prism [uncoated N-BK7 right angle prism], detector [New Focus Model-1811]. (b): Plot of the relative detected optical intensity, with the EOM used to shutter the optical field at $\sim175$~ns. The solid grey curve is an exponential  fit to the data over the range 221--454~ns. } \label{rdpic}
\end{figure}

The optical quality factor of the WGM resonator was characterized via cavity ringdown measurement~\cite{Vahala}, using the setup shown in Fig.~\ref{rdpic}(a). A fiber laser of wavelength $\lambda = 1550$~nm was critically coupled into the resonator using a prism mounted on a 3-axis nanopositioning stage. An optical intensity modulator was used to rapidly switch off the laser intensity. The exponential decay of light out of the resonator was then detected using a fast photodiode. The resulting cavity ringdown measurement is shown in Fig.~\ref{rdpic}(b). The cavity lifetime $\tau_e$ is determined to be 233~ns from an exponential fit to the data (grey line in Fig.~\ref{rdpic}(b)), which corresponds to an intrinsic optical quality factor of $Q \equiv \Omega \, \tau_e = 2\pi c \, \tau_e/\lambda=2.8\times 10^8$, where $\Omega$ is the angular frequency of the laser, and $c$ is the speed of light in vacuum~\cite{ringdown} .

\begin{figure}
\includegraphics[width=8.5cm]{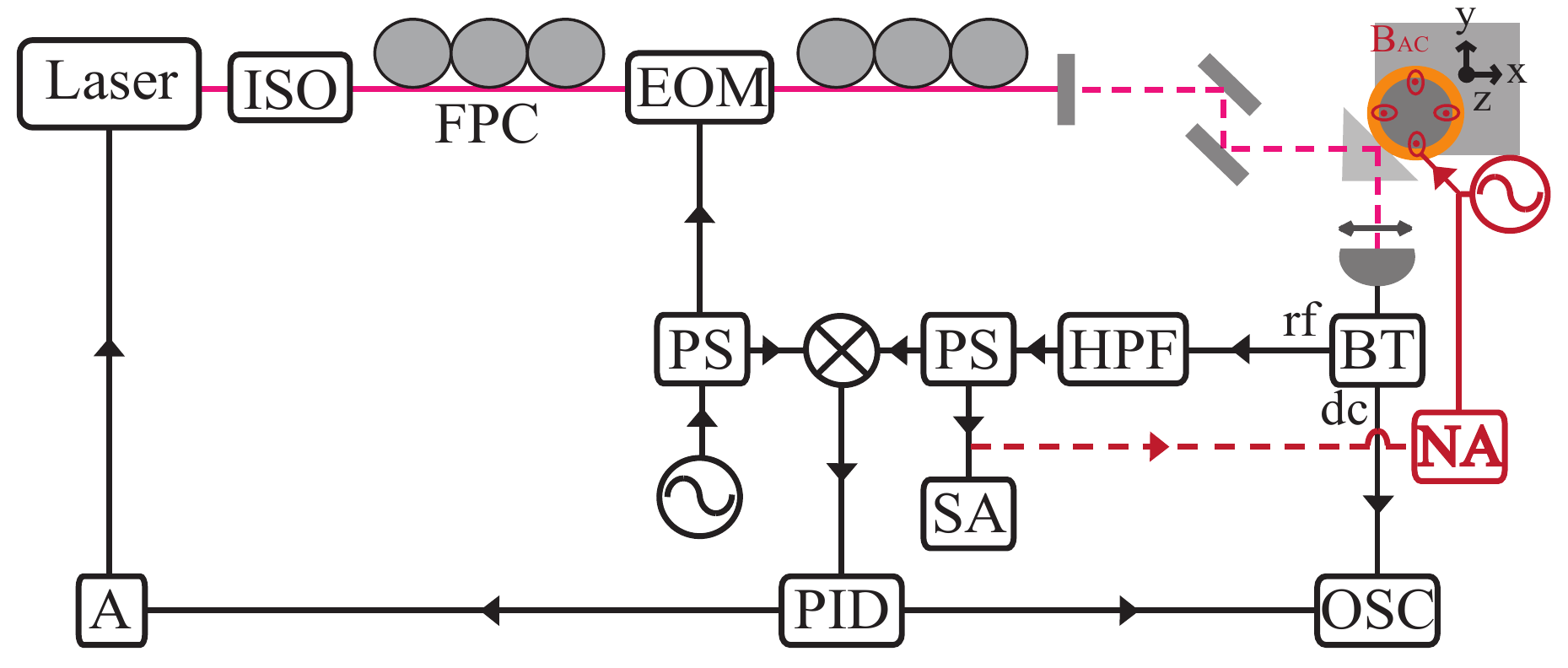}
\caption{\label{experiment}(Color online) A schematic of the experimental set-up used to perform magnetic field sensing. Laser [Koheras Adjustik C15], ISO: isolator [Thorlabs-OFR], EOM: electro-optic modulator [Covega Phase Modulator], NA: network analyzer [Agilent E5061B], SA: spectrum analyzer [Agilent N9010A], PID: proportional integral derivative controller [New Focus LB1005], BT: bias tee [Mini circuits 0.1-4200 MHz], HPF: high-pass filter [Mini circuits 0.07-1000 MHz], LPF: low pass filter [Mini circuits DC-1.9 MHz], PS: power splitter [Mini circuits 1-650 MHz], A: amplifier [ZFL-500].
} \label{experiment}
\end{figure}
\section{Experiment}
Fig.~\ref{experiment} shows a schematic of the measurement setup. Light from the fiber laser was passed through an isolator and an electro-optic modulator (EOM), and then evanescently coupled to the resonator in the same manner as described previously. The EOM was used to phase modulate the light at 13.6 MHz, well outside the resonator's linewidth ($\kappa/2\pi = \tau_e^{-1}/2\pi < 1$~MHz). The output field from the resonator was detected on an InGaAs photoreceiver. Electronic mixing of this output with a 13.6 MHz local oscillator generated a Pound-Drever-Hall (PDH) error signal~\cite{PDH}. This error signal provided a measure of the deviation of the laser frequency from the cavity resonance frequency. In a similar approach to Ref.~\cite{sensor2},
this signal was used both to lock the laser to the cavity resonance, and to detect the effect of applied magnetic fields on the length of the cavity -- i.e., it provided the magnetic field signal. To maximise the signal-to-noise ratio (SNR) of the sensor, a large modulation was applied to the EOM, transferring approximately half of the optical power into 13.6~MHz sidebands. It was found that only 40 $\mu$W of off-resonant light was required at the photoreceiver to resolve the noise of the optical field over the photoreceiver electronic noise floor.
A coil with diameter of 6.5 cm and a total of 60 turns was positioned above the resonator, and used to generate the signal magnetic field to be detected. The strength of this field was calibrated using a commercial Hall probe [Hirst GM04]. A neodymium magnetic was placed in close proximity to the resonator  to pre-polarize the Terfenol-D, thereby enhancing its linear response to applied magnetic fields~\cite{DCenhance,AM}. 

\section{Results and discussion}
\begin{figure}
\centering
\includegraphics[width=8cm]{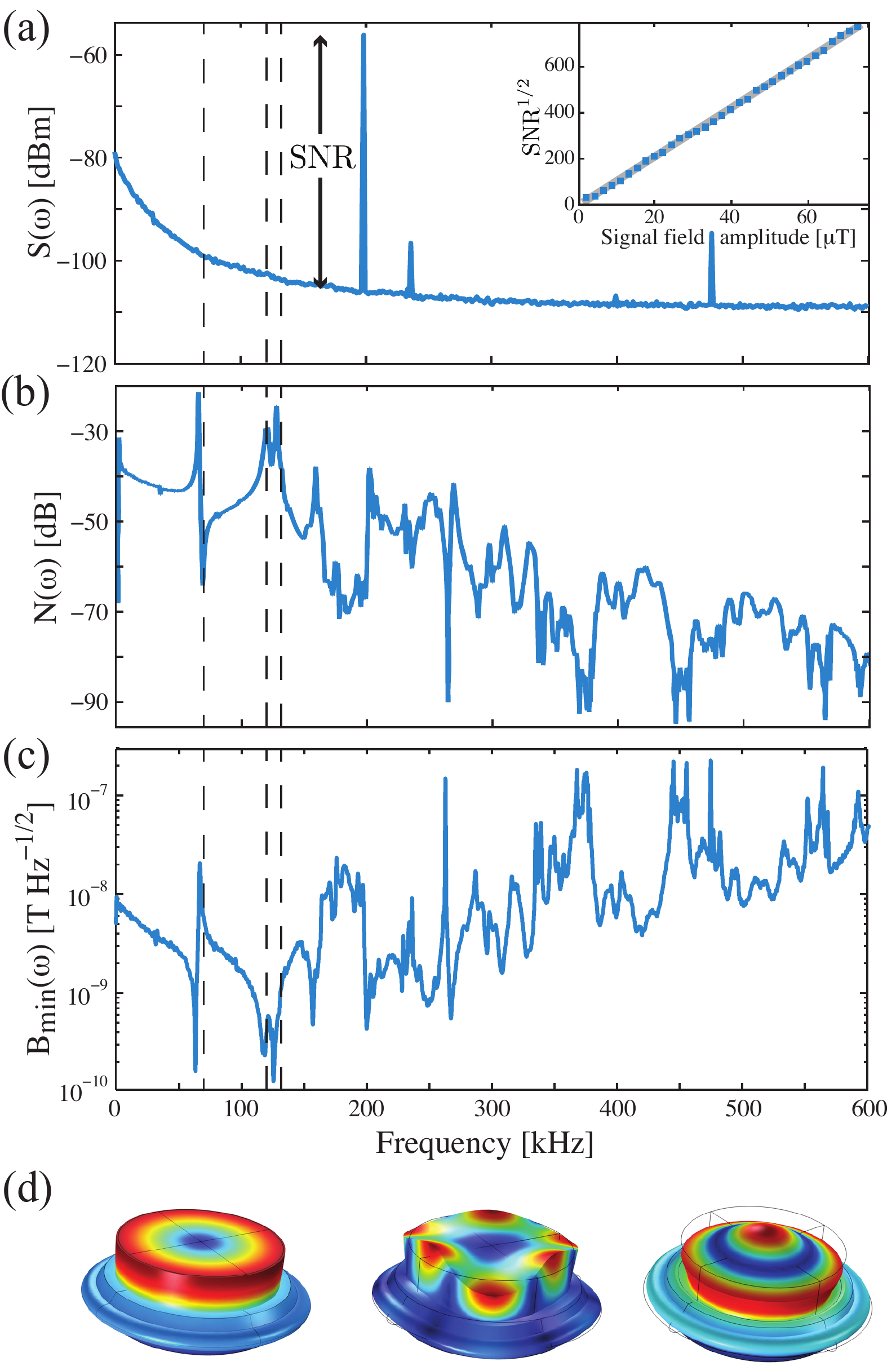}
\caption{(Color online) Experimental results. (a) Power spectral density $S(\omega)$ of the error signal at offset frequencies above the 
13.6~MHz optical sideband frequency, showing the response an applied magnetic field at 200 kHz. Inset: response to the magnetic field as a function of signal field strength, with 330 Hz spectrum analyzer resolution bandwidth. (b) System response $N(\omega)$ measured via network analysis as a function of applied magnetic field frequency. (c) Magnetic field sensitivity $B_{\rm min}(\omega)$ as a function of frequency. (d) Finite element modelling of mechanical eigenmodes of the device. From left to right, the modes are: the fundamental radial breathing mode at 69.8~kHz, a crown mode at 120.4~kHz, and the second order radial breathing mode at 131.9~kHz. The vertical dashed lines in (a)--(c) show the frequencies of these three modes.}\label{result}
\end{figure}

The response of the magnetometer to applied signal fields was characterised via spectral and network analysis of the PDH error signal. Fig.~\ref{result}a shows the power spectral density $S(\omega)$ of this error signal at frequencies above the 13.6 MHz optical sideband frequency, measured using a spectrum analyzer. It was verified that the resonator was capable of detecting magnetic fields by applying a reference magnetic field with root mean square (RMS) amplitude $B_{\rm ref}=7.8~\mu$T and frequency $\omega_{\rm ref}=$200 kHz. This caused a corresponding tone at 200~kHz in the power spectral density of the error signal  (see  Fig.~\ref{result}(a)). The magnetic field sensitivity at 200 kHz was then determined following Ref.~\cite{PRL}, as 
  \begin{equation}
  B_{\rm min}(\omega_{\rm ref})=\frac{B_{\rm ref}}{\sqrt{{\rm SNR} \times {\rm BW}}} = 1.4~\text{nT~Hz}^{-1/2},
  \end{equation}
 where ${\rm SNR}=49.7$~dB is the ratio of the signal height at $\omega_{\rm ref}$ to the corresponding noise floor (see Fig.~\ref{result}(a)), and ${\rm BW}=330$ Hz is the spectrum analyzer resolution bandwidth.  The dynamic range of the magnetometer was tested by measuring the response as a function of signal field amplitude. 
A linear response was observed over the full accessible range of signal field strengths, up to field strengths as large as 72 microtesla which exceeds the earth's field (see inset in Fig.~\ref{result}(a)). 

The spectrum analyser noise floor in Fig.~\ref{result}(a), combined with the system response, as quantified by network analysis, allowed the magnetic field sensitivity to be determined over the full hertz-to-kilohertz frequency range. Specifically, the magnetic field sensitivity is given by~\cite{PRL}
\begin{equation}
B_{\rm min}(\omega)=\sqrt{\frac{S(\omega)N(\omega_{\rm ref})}{S(\omega_{\rm ref})N(\omega)}}B_{\rm min}(\omega_{\rm ref}),
\end{equation}
where $S(\omega)$ is the noise power spectrum observed without any applied magnetic field, and $N(\omega)$ is the system response obtained by sweeping the frequency of the magnetic field and recording the power contained within the spectral peak using a network analyzer, shown in Fig.~\ref{result}(b). Below 140~kHz the structure in the system response is dominated by the three of mechanical eigenmodes of the device. Finite element simulations of these modes are shown in Fig.~\ref{result}(d), with the simulated frequencies matching closely to the observed frequencies evident in Fig.~\ref{result}(b). Note that the dispersive feature at the fundamental radial breathing mode resonance frequency ($69.8$~kHz) results from interference of the response of that mode and the background response of the device. Inspection of the measured error signal power spectrum (Fig.~\ref{result}(a)) shows that the thermomechanical noise of all of these three mechanical eigenmodes is beneath the laser phase noise floor, indicating that the precision of magnetic field measurement with this device will be limited by laser noise rather than thermomechanical noise. Above 140 kHz, the system response is suppressed with increasing frequency, with complex structure existing due to the presence of multifold higher frequency mechanical resonances.

Fig.~\ref{result}(c) shows the sensitivity measured over the frequency range from 127~Hz to 600~kHz. A peak sensitivity of 131~pT~Hz$^{-1/2}$ is achieved at 126.75 kHz, close to the eigenfrequencies of the mechanical crown and second order radial breathing modes, while similar sensitivity is also achievable at frequencies close to the fundamental radial breathing mode. Evidently, the sensitivity is enhanced by these mechanical resonances, and outperforms previous cavity optomechanical magnetometers in the same frequency range by around three orders-of-magnitude.  The best previously reported result had sensitivity above 130~nT~Hz$^{-1/2}$ over the full range of the measurements we report here~\cite{AM}.

The sensitivity of our current device is limited by laser phase noise at frequencies below 540 kHz and shot noise above that frequency. Consequently, improved sensitivity could be achieved using phase stabilization~\cite{nanoparticledetection}, increased optical power, or higher optical quality factors, until eventually the thermomechanical noise floor is reached~\cite{PRL,AM}. Quality factors as high as $Q =3\times10^{11}$ have been realized for millimeter-scale CaF$_2$ WGM resonator at 1550 nm and at room temperature~\cite{experimenthighQ}. Our sensitivity therefore could be further enhanced by fabricating a higher $Q$ resonator. The sensitivity could also be enhanced by engineering the structure of the device for improved overlap between the motion of mechanical eigenmodes and the stress applied by the Terfenol-D~\cite{theorysensitivity}. Estimates based on Ref.~\cite{model} indicate that sensitivities in the range of femtotesla may be obtainable with a full optimization.


\begin{acknowledgments}
The authors acknowledge valuable advice from Beibei Li, Glen Harris, George Brawley and Michael Taylor. This research was funded by the Australian Research Council Centre of Excellences CE110001013 and CE110001027, the Discovery Project DP140100734, and by DARPA via a grant through the ARO. Device fabrication was performed at the Australian National University. Changqiu Yu acknowledges support by the China Scholarship Council (File Number: LJF[2013]3009). WPB and PKL are supported by the ARC Future and Laureate Fellowship FT140100650 and FL150100019, respectively. 
\end{acknowledgments}

\end{document}